\documentclass[11pt]{article}

\usepackage{graphicx,amsmath,amsfonts,amssymb,dcolumn,amsthm,slashed,bbm,xspace,pgffor,tikz,mathbbol,latexsym,enumerate}
\usepackage{soul}
\usepackage[colorlinks,citecolor=red,urlcolor=cyan,linkcolor=blue]{hyperref}
\usepackage[utf8]{inputenc}
\usepackage[english]{babel}
\usepackage{lmodern}
\usepackage{microtype}
\usepackage{cite}

\numberwithin{equation}{section}

\begin{document}

\title{\textbf{Photon propagation in a material medium on a curved spacetime}}

\author{\textbf{Amanda Guerrieri}\thanks{amguerrieri@cbpf.br}\ \;and \textbf{Mário Novello}\thanks{mnovello42@gmail.com} \\\\
\textit{{\small CBPF - Centro Brasileiro de Pesquisas Físicas}}\\
\textit{{\small R. Dr. Xavier Sigaud, 150 - Urca, Rio de Janeiro - RJ, 22290-180}}}

\date{}
\maketitle

\begin{abstract}
We consider a nonlinear dielectric medium surrounding a static, charged and spherically symmetric compact body which gravitational field is driven by General Relativity (GR). Considering the propagating waves on the dielectric medium, we describe the trajectory of light as geodesics on an effective geometry given by Hadamard’s discontinuities. We analyze some consequences of the effective geometry in the propagation of light, with relation to the predictions of the background gravitational field, that includes corrections on the geometrical redshift and on the gravitational deflection of light. We show that the background electromagnetic field polarize the material medium, such that different polarizations of light are distinguished by different corrections on these quantities. As a consequence, we have two possible paths for the trajectory of light in such configuration, that coincide if we turn off the electromagnetic field or if the permittivity is constant. We show that the effective metric associated to the negative polarization, for a given dependence of the dielectric permittivity, is conformally flat.  
\end{abstract}

\newpage

\tableofcontents

\newpage

\section{Introduction}

Dielectric mediums have been extensively studied in literature, both theoretically \cite{lan84,T1,1984MolPh..52.1241G,1977UsFiN.123..349B,Jones1948ANC} and experimentally \cite{PhysRevLett.85.4478,PhysRevLett.88.063001,LlDXLAN,exp1}. In the past years, analogue models of gravity\cite{bookex,article,article1,article2,article3,article4,article5,article6,article7,article8, article9} considering a nonlinear dielectric medium received attention in the scientific community\cite{Novello2003EffectiveG,PhysRevD.78.045015,LK,LM,LKEM,PhysRevE.82.036605,LK2,LK3,LS,DK,LKR,Novello2000GeometricalAO,Novello:2001gk,N,Lorenci2012TrirefringenceIN,L,LRY, ME1, MSJVR}. These models predicted that nonlinear materials can produce, in laboratory, an effective (optical) horizon with similar properties to the event horizon of black holes\cite{ABH}. At these horizons, light is trapped, such that a ring could be observed due to birefringence phenomenon\cite{article10,article11,article12,article13}. Mathematically, this is described by an effective geometry \cite{Novello2003EffectiveG,LM,Novello:2001gk,N,L,ME1} generated by the medium, such that light follows geodesics in this effective geometry, and it determines the position of the optical horizons. In this paper, we are interested in applying this method to curved spacetimes. For that, instead of considering the flat spacetime of a laboratory as the background geometry, we are going to consider a curved background generated by a spherically symmetric object. This choice is motivated by studies concerning a system formed by black holes surrounded by a plasma \cite{osti_4024998,Perlick:2015vta}.  If this medium has nonlinear properties, we can describe its effects on the propagation of light by means of an effective geometry, in a similar manner to analogue models.

In order to apply this method to a system formed by a compact object surrounded by a nonlinear electromagnetic material, we are going to consider a dielectric medium $\mu=const.$, $\epsilon=\epsilon(E)$ with negligible mass, compared to the compact object, such that we can idealize a fixed background. In this paper, we show that this simple model is enough to predict some important consequences in the propagation of light, that includes corrections on the geometrical redshift and on the light deflection by a compact body. If the medium surround a charged body, we can have two polarization modes for light, such that this model gives a mechanism to evaluate the consequences of birefringence effects on curved spacetimes. Each mode is associated to a different correction on these quantities, such that we can have two possible paths for the light ray. These paths only coincide if the electrical field is turned off or if the permittivity is constant. The vacuum appears as a particular case in this configuration.

In this paper, what we call event horizon depends only on the background metric, and what we call killing horizon depends only on the effective metric, such that we are going to refer to it as an optical horizon. Formally, only in the vacuum we could use the background metric to evaluate the optical horizon, because in that case the effective metric and the background metric are equivalent. This is a consequence of the definition of a killing horizon \cite{Wald:1984rg,Carroll:2004st}, which is described as the radius where the redshift diverges and the time killing vector changes its sign, implying that we need to consider the effective metric to determine it. We proved that the position of the optical horizons coincide with the position of the killing horizons predicted by the background metric. This is a direct consequence of the form of the effective metric.

The paper is organized as follows: In section \ref{sec2} we review Hadamard’s method of discontinuities and construct the effective geometry for a nonlinear dielectric medium, in order to discuss birefringence effects on flat spacetimes. In section \ref{sec3}, we apply this method to a spherically symmetric curved background and show its consequences on the propagation of light. In section \ref{sec4}, we consider three particular cases for our analyzes. The first, is a special case that doesn't produce alterations on either the geometrical redshift or the gravitational deflection of light. However, it changes the effective potential of light, which coincides with the vacuum predictions only if $\epsilon=\epsilon_0$. The second, is a less restrict case that produce alterations on the effective potential and on the deflection angle of light, but maintain the predictions for the geometrical redshift of the positive polarization. The last, is a general case that account all possible consequences on curved spacetimes. Both can be reduced to the first case in the particular situation where the permittivity is constant.  Finally, we end our analyzes showing that the effective metric associated to the negative polarization can be written in a conformally flat form, for a given dependence of the dielectric permittivity. In Section \ref{FINAL}, we display our conclusions. 

In this article, we consider the notation for the partial derivatives $\bar{A}=\frac{\partial A}{\partial E}$ and $A' = \frac{\partial A}{\partial H}$. A Minkowskian spacetime is used in subsection \ref{sec.2} as the background metric $\gamma^{\mu\nu}$ and it has signature $(+, -, -, -)$. All quantities are refereed as measured by the observer $v^\mu$. In particular, we consider the definition $\Dot{X}^\mu=X^\mu_{;\nu} v^\nu$. For an arbitrary vector $X^\mu=(0, \vec{X})$ we define its modulus by considering the relation $X= (-X^\mu X_\mu)^{1/2}$ and its associated unit vector as $\hat{X}=\vec{X} / X$. For any two quantities X and Y we denote
its scalar product $X^\mu Y_\mu$ following the notation $(X.Y)$. The projection tensor is defined as $h^{\mu\nu} = g^{\mu\nu} - v^\mu v^\nu$ and Kronecker tensor is represented by $\delta^\mu_\nu$. Additionally, we consider Levi-civita tensor to be defined as $\eta^{\lambda\gamma\beta\sigma}= - \frac{1}{\sqrt{-g}} \epsilon^{\lambda\gamma\beta\sigma}$, where $g$ is the determinant of the background metric.

\section{Effective geometry}\label{sec2}
In order to find an effective geometry associated with the propagation of light inside a dielectric medium, we are going to consider Hadamard method \cite{Hadamard,Papapetrou:1975kj}. Let us define a surface of discontinuity $\Sigma (x^\mu)= const.$ which delimit locally two regions of the spacetime, represented by 1 and 2. Given a function $f$, we call $f^{(1)}$ and $f^{(2)}$ the values taken by the function on each domain. Hadamard's discontinuity of the function $f$, with relation to the surface $\Sigma$, is defined as
\begin{equation}
    [f(x)]\; \vert_{\Sigma} = \lim_{\epsilon \to 0^{+}} \left( f^{(1)}(x+\epsilon) - f^{(2)}(x-\epsilon)\right)\;,
\end{equation}
such that the point $x$ belongs to the surface. We suppose that $f$ is continuous on the surface $\Sigma$, but it's first derivatives $f_{,\alpha}$ don't
\begin{eqnarray}
\left[f\right] \; \vert_{\Sigma} &=& 0\;,\\
\left[f_{,\alpha}\right] \; \vert_{\Sigma} &\neq& 0 \;.
\end{eqnarray}
Considering the differentials of the function in both regions and knowing that the shift vector $dx^{\alpha}$ belongs to the surface $\Sigma$, we find
\begin{eqnarray}
df^{(i)} = \partial_{\alpha} f^{(i)} dx^{\alpha}\;,
\end{eqnarray}
with $i=\{1,2\}$ representing both regions. Hadamard showed that these differentials should exist and be continuous on the surface, i.e., $[df] \vert_{\Sigma}=0$. That is, the discontinuity of the derivatives of $f$ must be an object orthogonal to the surface, as a consequence of
\begin{equation}
    [df] \vert_{\Sigma}=[\partial_\alpha f] \vert_{\Sigma} \;dx^{\alpha}=0.
\end{equation}
Thus, there exists a scalar $\sigma(x)\neq 0$ such that $[f_{,\alpha}]\;\vert_{\Sigma}=\sigma(x) \; k_\alpha$ with $k_\alpha=\Sigma_{,\alpha}$. In the following subsection, we are going to apply this method to the equations of electrodynamics on a material medium.

\subsection{Propagating waves on a dielectric medium}\label{sec.2}
 The propagation of light inside a nonlinear medium is described by null geodesics in an effective geometry, represented by an effective metric. This effective metric yields modifications on the background metric, associated to the functions of the medium. The constitutive relations of a material medium are given by
\begin{eqnarray}
D_{\alpha} &=& \epsilon_\alpha^\beta (E^\mu,H^\mu) E_\beta\label{13}\\
B_{\alpha} &=& \mu_\alpha^\beta (E^\mu,H^\mu) H_\beta \;,\label{14}
\end{eqnarray}
where $\epsilon_\alpha^\beta$ and $\mu_\alpha^\beta$ represents the permittivity and the permeability tensors of the medium. They relate the electric field $E^\mu$ and the magnectic field $B^\mu$ with the displacement $D^\mu$ and the auxiliary $H^\mu$ fields. We can decompose the tensors that represent the electromagnetic field, its dual and the polarization field, in terms of an observer's field $v^\mu$, as
\begin{eqnarray}
F_{\mu\nu} &=& E_\mu v_\nu - E_\nu v_\mu + \eta_{\mu\nu}^{\phantom{\mu\nu}\rho\sigma}v_\rho B_\sigma\\
F_{\mu\nu}^* &=& B_\mu v_\nu - B_\nu v_\mu - \eta_{\mu\nu}^{\phantom{\mu\nu}\rho\sigma}v_\rho E_\sigma\label{dual}\\
P_{\mu\nu} &=& D_\mu v_\nu - D_\nu v_\mu + \eta_{\mu\nu}^{\phantom{\mu\nu}\rho\sigma}v_\rho H_\sigma\;\label{Fmat},
\end{eqnarray}
in order to apply it to Maxwell's equations in the absence of sources $^* F^{\mu\nu}_{\phantom{\mu\nu};\nu}=0$ and $P^{\mu\nu}_{\phantom{\mu\nu};\nu}=0$. For an isotropic medium, one is able to write
\begin{eqnarray}
       \epsilon^{\beta}_{\alpha}(E^\mu,H^\mu) &=& \epsilon(E,H) (\delta^{\beta}_{\alpha} - v^\beta v_\alpha)\nonumber\\
       \mu^{\beta}_{\alpha}(E^\mu,H^\mu) &=& \mu(E,H) (\delta^{\beta}_{\alpha} - v^\beta v_\alpha)\;,\label{211}
\end{eqnarray}
where $E$ and $H$ are the modulus of the electromagnetic field. By considering these relations, Maxwell's equations become  
\begin{eqnarray}
- \epsilon\;E^\mu \Dot{v_\mu}-\epsilon \;E^{\alpha}_{\phantom{\alpha},\alpha} + E^{\alpha} \left(\frac{\bar{\epsilon}}{E}E^\lambda E_{\lambda , \alpha} + \frac{\epsilon'}{H}H^\lambda H_{\lambda , \alpha}\right) + \eta^{\mu\nu\rho\sigma}v_{\rho,\nu}v_\mu H_\sigma  &=&  0\label{max1}\\
 - \mu\;H^\mu \Dot{v_\mu}-\mu\; H^{\alpha}_{\phantom{\alpha},\alpha} + H^\alpha \left(\frac{\bar{\mu}}{E}E^\lambda E_{\lambda , \alpha} + \frac{\mu'}{H}H^\lambda H_{\lambda , \alpha}\right) - \eta^{\mu\nu\rho\sigma}v_{\rho,\nu}v_\mu E_\sigma &=&  0 \\
\epsilon\;( \Dot{E}^\lambda +  E^\lambda v^\nu_{,\nu} -  E^\nu v^\lambda_{,\nu}) - E^\lambda \left( \frac{\bar{\epsilon}}{E}E^\lambda E_{\lambda , \alpha} + \frac{\epsilon'}{H}H^\lambda H_{\lambda , \alpha}\right)v^\alpha + \eta^{\lambda\nu\rho\sigma}(v_{\rho,\nu} H_\sigma + v_\rho H_{\sigma,\nu}) &=&  0\\
\mu\;( \Dot{H}^\lambda + \;H^\lambda v^\nu_{,\nu}  -  H^\nu v^\lambda_{,\nu}) - H^\lambda \left( \frac{\bar{\mu}}{E}E^\lambda E_{\lambda , \alpha} + \frac{\mu'}{H}H^\lambda H_{\lambda , \alpha} \right) v^\alpha - \eta^{\lambda\nu\rho\sigma} (v_{\rho,\nu} E_\sigma + v_\rho E_{\sigma,\nu}) &=&  0\;.\label{max4}
\end{eqnarray}
Hence, we can consider the definition of Hadamard's discontinuities. It gives
\begin{eqnarray}
\left[E_\mu\right]_{\Sigma}=0\;,\;\;\;\;\; \left[E_{\mu,\lambda}\right]_{\Sigma} &=& e_\mu k_\lambda \;,\nonumber\\
\left[H_\mu\right]_{\Sigma}=0\;,\;\;\;\;\;\left[H_{\mu,\lambda}\right]_{\Sigma} &=& h_\mu k_\lambda\;,\nonumber
\end{eqnarray}
where $e^\mu$ and $h^\mu$ represent the discontinuities of the fields on the surface $\Sigma$ and $k_\lambda$ is the wave 4-vector. When we apply the discontinuities to Maxwell's equations (\ref{max1}-\ref{max4}) the terms proportional to $v_{\mu,\nu}$ disappear, since they are proportional to the discontinuities of the electromagnetic field. Therefore, for an arbitrary observer, we have
\begin{eqnarray}
\label{sis1}\epsilon \;e^\alpha k_\alpha - E^{\alpha} \left(\frac{\bar{\epsilon}}{E}E^\lambda e_\lambda k_\alpha + \frac{\epsilon'}{H}H^\lambda h_\lambda k_\alpha \right) &=&  0\\
 \mu\; h^{\alpha}k_{\alpha} - H^\alpha \left(\frac{\bar{\mu}}{E}E^\lambda e_\lambda k_\alpha + \frac{\mu'}{H}H^\lambda h_\lambda k_\alpha\right) &=&  0 \\
\epsilon\; e^\lambda k_\alpha v^\alpha - E^\lambda \left(\frac{\bar{\epsilon}}{E}E^\beta e_\beta k_\alpha v^\alpha + \frac{\epsilon'}{H}H^\beta h_\beta k_\alpha v^\alpha \right) + \eta^{\lambda\nu\rho\sigma}  v_\rho h_{\sigma}k_{\nu} &=&  0\\
\mu\; h^\lambda k_\alpha v^\alpha - H^\lambda \left( \frac{\bar{\mu}}{E}E^\beta e_\beta k_\alpha v^\alpha + \frac{\mu'}{H}H^\beta h_\beta k_\alpha v^\alpha\right) - \eta^{\lambda\nu\rho\sigma}  v_\rho e_{\sigma}k_{\nu} &=&  0\;,\label{sis2}
\end{eqnarray}
where $\Dot{A}^\mu=A^\mu_{,\nu} v^\nu$ was written in terms of its discontinuities as $[\Dot{A}^\mu]_\Sigma= a^\mu k_\nu v^\nu$. In order to solve this system of equations and find the effective geometry associated to the propagation of light inside a dielectric medium, we will set the magnetic permeability $\mu=const.$ and the electric permittivity  $\epsilon=\epsilon(E)$. These relations simplify the system of equations, which becomes
\begin{eqnarray}
\epsilon \;e^\alpha k_\alpha - \frac{\bar{\epsilon}}{E}E^\lambda e_\lambda E^{\alpha} k_\alpha  &=&  0\label{die1}\\
 \mu\;  h^{\alpha}k_{\alpha} &=&  0 \label{die2}\\
\epsilon\; e^\lambda k_\alpha v^\alpha - \frac{\bar{\epsilon}}{E}E^\beta e_\beta k_\alpha v^\alpha E^\lambda  + \eta^{\lambda\nu\rho\sigma}  v_\rho h_{\sigma}k_{\nu} &=&  0\label{die3}\\
\mu\; h^\lambda k_\alpha v^\alpha  - \eta^{\lambda\nu\rho\sigma}  v_\rho e_{\sigma}k_{\nu} &=&  0\label{die4}\;.
\end{eqnarray}
This system can be solved and its solution is extensively known in literature \cite{Novello2003EffectiveG,PhysRevD.78.045015,LK,LM,LKEM,PhysRevE.82.036605,LK2,LK3,LS,DK,LKR,Novello2000GeometricalAO,Novello:2001gk,N,Lorenci2012TrirefringenceIN,L,LRY, ME1, MSJVR}. It is direct to see that when we contract the last two equations with $k_\lambda$ we obtain the first two equations. Therefore, it is sufficient to work only with equations (\ref{die3}-\ref{die4}). The last equation permits us to isolate
\begin{eqnarray}
h_\sigma = \frac{1}{\mu k^\alpha v_\alpha}\eta_{\sigma\nu\rho\mu}v^\rho e^\mu k^\nu \;.\label{rel1}
\end{eqnarray}
By considering the contraction of
\begin{eqnarray}
\eta^{\lambda\gamma\beta\sigma}\eta_{\sigma\nu\rho\mu}v^\rho e^\mu k^\nu k_\gamma v_\beta  &=& k^\lambda\left( v^\rho k_\rho e^\mu v_\mu - e^\mu k_\mu\right) + v^\lambda \left(  e^\mu k_\mu k^\nu v_\nu - e^\mu v_\mu k^\nu k_\nu \right) +\nonumber\\
&+& e^\lambda \left(k^\nu k_\nu  - v^\rho k_\rho  k^\nu v_\nu  \right)\;,
\end{eqnarray}
we can substitute relation \eqref{rel1} in equation \eqref{die3}. One is able to notice that relations $F_{\mu\nu} v^\mu v^\nu = F_{\mu\nu}^* v^\mu v^\nu=0$ imply that conditions $E_\mu v^\mu  = B_\mu v^\mu =0$ are valid. Therefore, they can be written in terms of their discontinuities as $e^\mu v_\mu = h^\mu v_\mu = 0$. Thus, simplifying \eqref{die3}, we can obtain Fresnel tensor $Z^{\mu\lambda}$ by isolating a relation $Z^{\mu\lambda} e_\lambda=0$, given by
\begin{eqnarray}
Z^{\mu\lambda} &=&(k. v)v^\mu k^\lambda  - (k.v)^2\;\mu \frac{\bar{\epsilon}}{E}E^\lambda E^\mu + \gamma^{\mu\lambda} \left[ (\mu\epsilon - 1) (k.v)^2 +  k^\nu k_\nu \right] - k^\mu k^\lambda \;.
\end{eqnarray}
Condition det$[Z^{\mu\lambda}]=0$ generalizes Fresnel equation, such that non-trivial solutions can be found. One way to solve the above equation is to expand $e_\lambda$ as a linear combination of four linearly independent vectors, which can be choosen as $E_\lambda$, $\eta_{\nu\rho\mu\lambda}v^\nu E^\rho k^\mu$, $k_\lambda$, $v_\lambda$ such that $ e_\lambda = \alpha_0 E_\lambda + \beta_0\;  \eta_{\nu\rho\mu\lambda}v^\nu E^\rho k^\mu + \gamma_0 k_\lambda + \delta_0 v_\lambda$. It follows that the condition $Z^{\mu\lambda} e_\lambda=0$ can be rewritten as
\begin{eqnarray}
0&=& \alpha_0 \left[\frac{k^\nu k_\nu}{(k.v)^2}-1 + \mu \frac{\partial (\epsilon E)}{\partial E} \right]  - \gamma_0 \left[\frac{\mu\bar{\epsilon}}{E}\; E^\alpha k_\alpha\right] \\
\nonumber\\
0&=& \beta_0 \left[ (\mu\epsilon - 1) (k.v)^2 +  k^\nu k_\nu \right] \label{88}\\
0&=& -k^\lambda (\alpha_0 E_\lambda  + \delta_0 v_\lambda) - \gamma_0 (1-\mu\epsilon)(k.v)^2\\
0&=& (k.v)k^\lambda (\alpha_0 E_\lambda + \gamma_0 k_\lambda)+ \delta_0 \left[ \mu\epsilon\; (k.v)^2 +  k^\nu k_\nu \right]\;.\label{sist}
\end{eqnarray}
This system of equations can be easily solved. It follows the relations
\begin{eqnarray}
    \delta_0 &=& - \gamma_0 (k.v) \nonumber\\
    \alpha_0 &=& \frac{\mu\epsilon (k.v)^2}{E^\lambda k_\lambda} \gamma_0\;, 
\end{eqnarray}
such that one is able to find two solutions, respectively,
\begin{eqnarray}
  k^\mu k_\mu&=&  [ 1-\epsilon\;\mu - \bar{\epsilon}\;\mu E ] (k.v)^2 + \frac{\bar{\epsilon}}{\epsilon E}   E^\mu E^\nu k_\mu k_\nu\label{disp1}\\
 k^\mu k_\mu &=& [1-\mu\epsilon(E)](k.v)^2\label{disp2}
\end{eqnarray}
for the dispersive relation $\hat{g}^{\mu\nu}_\pm k_\mu k_\nu =0$ of each polarization mode
\begin{eqnarray}
       e_\lambda^+ &=& \rho^+ \{\mu\epsilon (k.v)^2 E_\lambda + E^\alpha k_\alpha [ k_\lambda - (k.v) v_\lambda]\} \nonumber\\
       e_\lambda^- &=& \rho^-\;  \eta_{\nu\rho\mu\lambda}v^\nu E^\rho k^\mu\;,
\end{eqnarray}
where $\rho^\pm$ are arbitrary constants. They can be described in terms of two effective metrics 
\begin{eqnarray}
    \hat{g}^{\mu\nu}_+ &=& \gamma^{\mu\nu} +  v^\mu\; v^\nu ( \epsilon\;\mu + \bar{\epsilon}\;\mu E -1) - \frac{\bar{\epsilon}}{\epsilon E}   E^\mu E^\nu\nonumber\\
    \hat{g}^{\mu\nu}_- &=& \gamma^{\mu\nu} +  v^\mu\; v^\nu ( \epsilon\;\mu -1)\;.\label{ef1}
\end{eqnarray}
These effective geometries describe the propagation of light inside a dielectric material medium with $\mu=const.$, $\epsilon=\epsilon(E)$. In the case where $\epsilon=const.$ the metrics are equal $\hat{g}^{\mu\nu}_\pm=\hat{g}^{\mu\nu}$ and the solution is given by Gordon metric \cite{gordon,NE},
\begin{equation}
       \hat{g}^{\mu\nu} = \gamma^{\mu\nu} + v^\mu\; v^\nu ( \mu\epsilon\; -1) \;.\label{Gordon}
\end{equation}
In this work, we are interested to analyze  dielectric mediums. The most general case \eqref{211} can not be described by an effective metric (see for instance \cite{PhysRevD.78.045015}), but there are other cases that by imposing some restrictions on the constitutive relations (\ref{13}-\ref{14}) permit us to find an effective metric, see \cite{LM,LK2,L,LKEM}. Some works extended this analysis to anisotropic material mediums \cite{LKR,LS,LM,LK2}, and showed that in some cases it is also possible to find an effective metric associated to the solution. 

 Given the definitions for the angular frequency of the electromagnetic wave $\omega= k^\alpha v_\alpha$ and the wave vector $q^\mu = h^\mu_{\phantom{\mu}\nu} k^\nu = k^\mu - \omega v^\mu$, one is able to rewrite $k^\mu=(\omega,\vec{q})$ such that $k^2 = \omega^2 - |\vec{q}|^2$ permit us to isolate the velocity of light in a material medium $v^2 = \omega^2 / |\vec{q}|^2$, associated to each polarization. Considering the dispersive relations (\ref{disp1}-\ref{disp2}) we find
\begin{eqnarray}
       v_+^2 &=& \frac{1}{\mu \partial(\epsilon E)/\partial E} \left[1 + \frac{E}{\epsilon}\frac{\partial \epsilon}{\partial E} (\hat{q}.\hat{E})^2\right] \nonumber\\
       v_-^2&=& \frac{1}{\mu \epsilon(E)}\;.
\end{eqnarray}
In the particular case where the propagation occurs on the direction of the electrical field $\hat{q}.\hat{E}=1$, both velocities are equal. This velocity is called velocity of the ordinary ray. Complementarly, one can define the maximum value of $v_+^2$ when $\hat{q}.\hat{E}=0$. If we set $c=1$, the index of refraction is defined as $v_\pm^{-1}$ such that the difference between the maximum and the minimum values of the index is given by 
\begin{equation}
    n_\perp - n_\parallel = \sqrt{\mu \partial(\epsilon E)/\partial E} - \sqrt{\mu \epsilon(E)} \;.
\end{equation}
In the literature \cite{1977UsFiN.123..349B,PhysRevLett.88.063001,exp1}, it was verified experimentally that permittivity $\epsilon(E)= \varepsilon + \alpha E^2$ results in the difference
\begin{equation}
     n_\perp - n_\parallel \approx \frac{\sqrt{\mu\varepsilon}\;\alpha}{\varepsilon}E^2\;.
\end{equation}
This situation describes a birefringence phenomenon known as Kerr effect. Another birefringence phenomenom well known in literature arises when one consider a permittivity $\epsilon(E)= \varepsilon + \beta E$, which produces the difference
\begin{equation}
     n_\perp - n_\parallel \approx \frac{\sqrt{\mu\varepsilon}\;\beta}{2\varepsilon}E\;,
\end{equation}
and it is associated to Pockels effect\cite{photonics}. We are interested to analyze the consequences of birefringence phenomenon in the propagation of light on a curved background, to evaluate what kind of changes it may cause in the predictions of General Relativity (GR). We want to know if a distant observer could infer if the medium has nonlinear electromagnetic properties only by measuring the geometrical redshift and the light deflection by a compact body. We expect to anwser if a correction on these quantities could lead an observer to determine the dependence $\epsilon(E)$ at distance, i.e, from a laboratory on earth.

\section{Dielectric medium on curved spacetimes}\label{sec3}
If we consider a dielectric medium with constant permeability $\mu$ and allow the permittivity to depend on the electric field $\epsilon(E)$, the propagation of light is given by geodesics in the effective metrics\footnote{Following Hadamard's method, one have that for an arbitrary background the discontinuities of the fields are represented by $\left[E_\mu\right]_{\Sigma}=0\;,\; \left[E_{\mu;\lambda}\right]_{\Sigma} = e_\mu k_\lambda$ and  $\left[H_\mu\right]_{\Sigma}=0\;,\;\left[H_{\mu;\lambda}\right]_{\Sigma} = h_\mu k_\lambda$. It then follows the same system of equations found in subsection \ref{sec.2}, only changing $\gamma^{\mu\nu}\rightarrow g^{\mu\nu}$. Since the terms proportional to $v^\mu_{; \nu}$ depend on the electromagnetic field, after taking their discontinuities ($[E_\mu]_\Sigma=0$, $[H_\mu]_\Sigma=0$) these terms vanish. For that reason, the system of equations and the results are valid for an arbitrary observer (and an arbitrary background metric).} 
\begin{eqnarray}
    \hat{g}^{\mu\nu}_+ &=& g^{\mu\nu} +  v^\mu\; v^\nu ( \epsilon\;\mu + \bar{\epsilon}\;\mu E -1) - \frac{\bar{\epsilon}}{\epsilon E}   E^\mu E^\nu\nonumber\\
    \hat{g}^{\mu\nu}_- &=& g^{\mu\nu} +  v^\mu\; v^\nu ( \epsilon\;\mu -1)\;,\label{ef}
\end{eqnarray}
where $g^{\mu\nu}$ is a metric that describes the background gravitational field. In this article, we will focus on static spherically symmetric solutions of GR. We set for a massive charged body 
\begin{equation}
     ds^2= \Delta(r)\; dt^2 - \Delta^{-1}(r) \;dr^2 -
    r^2 d\Omega^2 \;,\label{gnrN}
\end{equation}
where
\begin{eqnarray}
    \Delta&=& 1 + \frac{r_Q^2}{r^{2}} - \frac{r_h}{r}\;, \label{delta}
\end{eqnarray}
with the constants
\begin{equation}
    r_Q^2=\frac{\pi G}{c^4} \frac{Q^2}{4\pi^2 \epsilon_0}\;,\;\;\;\; r_h = \frac{2MG}{c^2}\;.
\end{equation}
In this work, we are going to consider the units $c=G=1$ and that the electromagnetic field, which appears in \eqref{ef}, is the electromagnetic field produced by the charged compact body.

At low energies, we can consider a fixed a background. Timelike particles follow geodesics on this background spacetime, while photons propagate following null geodesics in the effective geometry. This limit is sufficient to enumerate physical consequences of the theory, like changes on the redshift and on the gravitational deflection of light. By considering a killing horizon as a null surface where the redshift diverges and 
\begin{equation}
    \frac{dr}{dt} = \sqrt{\frac{\hat{g}_{00} (r)}{\hat{g}_{11} (r)}} = 0
\end{equation}
is satisfied, we are able to quantify possible changes in the location of these surfaces. Once the killing horizon depends only on the effective geometry, we call it an optical horizon.

Note that Einstein's equations are related to the background metric, not to the effective metric. So, all the discussion well known in literature about the location of the singularity at $r=0$ remains valid. The effective geometry will not change that, it can only alter definitions that depend directly on the propagation of light, as quoted above. 

The effective metrics \eqref{ef} have the inverse
\begin{eqnarray}
    \hat{g}_{\mu\nu}^+ &=& g_{\mu\nu} - (1-f_+)\;v_\mu v_\nu  + \frac{\xi}{1 + \xi} l_\mu l_\nu \nonumber\\
    \hat{g}_{\mu\nu}^- &=& g_{\mu\nu} - (1-f_-)\;v_\mu v_\nu \;, \label{cov}
\end{eqnarray}
where we define 
\begin{equation}
    f_- = \frac{1}{\mu \epsilon(E)}\;, \; \; \;f_+ = \frac{1}{\mu \epsilon(1 + \xi)}\;, \; \; \; \xi = \frac{\bar{\epsilon} E}{\epsilon}\;, \; \; \; \bar{\epsilon}= \frac{d\epsilon}{dE}\;, \; \; \; l_\mu = \frac{E_\mu}{E}\;.\label{defmeio}
\end{equation}
If we consider an observer in the dielectric comoving frame $v_\mu =  \sqrt{g_{00}}\delta_\mu^0$ and that the electric field $E(r)$ is radial and static $l_\mu = \sqrt{g_{11}}\delta^1_\mu$, we can simplify the metric components 
\begin{eqnarray}
    \hat{g}_{00}^+ &=&  \Delta f_+\nonumber\\
    \hat{g}_{11}^+ &=& -\frac{\Delta^{-1}}{1 + \xi}\;,\nonumber\\
     \label{comp3}\\
    \hat{g}_{00}^- &=&  \Delta f_-\nonumber\\
    \hat{g}_{11}^- &=&  -\Delta^{-1}\;, \nonumber
\end{eqnarray}
and the other components are the same as the background metric. It is clear they are equivalent when $\xi=0$, i.e., $\epsilon=const.$ . In that case, they reduce to Gordon metric and both polarizations follow the same path. Only in the vacuum the effective metrics are reduced to the background.

\subsection{Effective potential}
In order to obtain the paths followed by the photons, we vary the action
\begin{equation}
    \delta \int \left(\hat{g}_{00}\; \dot{t}^2 \;+\; \hat{g}_{11}\; \dot{r}^2 \;+\; \hat{g}_{22}\; \dot{\theta}^2 \;+\; \hat{g}_{33}\; \dot{\phi}^2\right)d\lambda =0\;
\end{equation}
and find two equations associated to the coordinates $t,\phi$. The equation for $\theta$ is identically satisfied if we choose $\dot{\theta}=0$ and $\theta=\pi /2$. This result implies in the conservation of total energy $W_0$ and the conservation of the angular momentum $L_0$,
\begin{eqnarray}
    \hat{g}_{00} \;\dot{t} &=& W_0\;,\nonumber\\
    r^2 \dot{\phi} &=& L_0\;.
\end{eqnarray}
For the variable $r$, we consider $ds^2=0$ in order to obtain
\begin{eqnarray}
    \dot{r}^2 &=& W_0^2 - V(r)\;,
\end{eqnarray}
with the effective potential $V(r)$ being defined as
\begin{equation}
  V(r)= W_0^2 -\frac{ L_0^2}{r^2 \;\hat{g}_{11}}\; + \frac{ W_0^2}{\hat{g}_{11}\;\hat{g}_{00}}\;.
\end{equation}
We can use the effective metrics \eqref{cov} to find an explicit form for the potentials, 
\begin{eqnarray}
    V_+(r)&=& \frac{ L_0^2 \Delta}{r^2} (1 + \xi)\; + \left[ 1 -  \frac{ (1 + \xi)}{f_+}\right]W_0^2\nonumber\\
    V_-(r)&=& \frac{ L_0^2 \Delta}{r^2}\; + \left[1-  \frac{ 1}{f_-}\right]W_0^2\;, \label{efpot}
\end{eqnarray}
which are valid for all $\epsilon(E)$. Note that, as expected, for the vacuum $f_\pm (r)=1$, $\xi = 0$ they are equal and can be reduced to Reissner-Nordstrom potential
\begin{eqnarray}
    V_{B}(r)&=& \frac{ L_0^2}{r^2} \left(1 + \frac{r_Q^2}{r^{2}} - \frac{2M}{r}\right)\; \nonumber\;,
\end{eqnarray}
which gives Schwarzschild potential when $r_Q=0$.  Particular cases will be analyzed in the next chapter.

\subsection{Gravitational deflection of light}

 The formula for the deflection of a light ray propagating from infinity to a point at a distance $r$ from the origin of a static and spherically symmetric compact body is known in literature\cite{Weinberg:1972kfs}. We can compute how it changes in the presence of a nonlinear dielectric medium. Generally, given this symmetry, the light deflection caused by a compact body surrounded by a dielectric medium is represented by
\begin{equation}
    \phi(r) - \phi_\infty = \int^{\infty}_{r} \sqrt{| \hat{g}_{11}(r)|}\left[\left(\frac{r}{r_0}\right)^2 \frac{\hat{g}_{00}(r_0)}{\hat{g}_{00}(r)} - 1\right]^{-1/2} \frac{dr}{r}\;,\nonumber
\end{equation}
where $\phi_\infty= \phi(\infty)$ and $r_0$ is the distance of maximum approximation, defined as the radius where the following relation is valid
\begin{equation*}
    \frac{dr}{d\phi} =0\;.
\end{equation*}
The angle of deflection in a region far from the source is given by\cite{Weinberg:1972kfs}
\begin{equation}
    \Delta \phi = 2 |\phi(r_0) - \phi_\infty| - \pi \;. \label{deflection}
\end{equation}
This angle predicts a correction measured by a distant observer.
So, for $\Delta_0=\Delta(r_0)$, the light deflection caused by a dielectric on a Reissner-Nordstrom background is
\begin{eqnarray}
       \phi^+(r) - \phi^+_\infty &=&
       \int^{\infty}_{r} \sqrt{ \frac{\Delta^{-1}}{1 + \xi} }\;.\;\left[\left(\frac{r}{r_0}\right)^2 \frac{\Delta_0 f_{+}(r_0)}{\Delta f_+} - 1\right]^{-1/2} \frac{dr}{r}\;,\label{gd}
\end{eqnarray}
\begin{eqnarray}
       \phi^-(r) - \phi^-_\infty &=&
       \int^{\infty}_{r} \sqrt{\Delta^{-1}}\;.\;\left[\left(\frac{r}{r_0}\right)^2 \frac{\Delta_0 f_{-}(r_0)}{\Delta f_-} - 1\right]^{-1/2} \frac{dr}{r}\;,\label{gd2}
\end{eqnarray}
for each polarization. We recall that relations \eqref{defmeio} are valid, i.e., $f_\pm=f_\pm (r),\;\xi=\xi(r)$ because $E=E(r)$. To compute this integral, it is necessary to define the electric permittivity  $\epsilon(E)$ of the material medium, and the electric field $E(r)$. In the trivial case where the electric field is constant, we have that the functions $f,\;\xi,\;\epsilon$ are going to be constants with relation to the variable of integration $r$.

\subsection{Geometrical redshift}
The geometrical redshift\cite{inverno:1992} is a phenomenon that permit electromagnetic waves traveling on a background gravitational field to lose energy by decreasing its wave frequency and increasing its wavelength. The geometrical redshift of a light ray emitted radially at $r_1$
 and detected at $r_2$
 with $r_1<r_2$, is given by
\begin{equation}
    1 + z = \frac{d \tau_2}{d \tau_1} = \sqrt{\frac{\hat{g}_{00} (r_2)}{\hat{g}_{00} (r_1)}}\;,
\end{equation}
where $d \tau_i$ defines the time separation between successive wave crests measured by the clocks. If $r_1=r$ and $r_2 \longrightarrow \infty$,  given $f^\pm_\infty = f_\pm(\infty)$ it follows the relation
\begin{equation}
      (1+z)_\pm = \sqrt{\frac{f^\pm_\infty}{\Delta f_\pm}}\;.
\end{equation}
Note that for the case $f_\pm=const.$ the redshift remains unchanged, with relation to the  predictions of the background 
\begin{equation}
      (1+z)_{B} = \sqrt{\frac{1 }{\Delta }}\;.\label{138}
\end{equation}
This case is given by a special class of materials, which includes the vacuum, and will be analyzed in the next chapter. For $f_\pm\neq const.$, we are considering a electromagnetic nonlinear medium on the background, defined by relations \eqref{defmeio}. It's appropriate to rewrite this general case as
\begin{equation}
      \frac{(1+z)_\pm}{(1+z)_{B}} =  \sqrt{\frac{f^\pm_\infty}{f_\pm}}\;.\label{139}
\end{equation}
We note that materials that satisfy 
\begin{equation}
    \epsilon(E) = \varepsilon + \sum_n c_n E^n\;, \label{mm}
\end{equation}
with $E(r)= r_Q/r^\alpha$ for $\alpha,\varepsilon, c_n>0$, have $f^\pm_\infty=const.$ and 
\begin{eqnarray}
       \sqrt{\frac{f^\pm_\infty}{f_\pm}} > 1\;,
\end{eqnarray}
This implies that the intensity of the redshift is stronger than predicted by the background gravitational field, for both polarizations. If $c_n<0$ it is clear that the opposite will occur (it have it intensity attenuated). The constant $\varepsilon \neq 0$, otherwise a null electrical field would imply in $\epsilon=0$, which isn't a physical situation. If we allow $\varepsilon < 0$, the above analyzes for materials \eqref{mm} with $c_n>0$ imply
\begin{eqnarray}
       \sqrt{\frac{f^\pm_\infty}{f_\pm}} < 1\;,
\end{eqnarray}
such that the redshift has it intensity reduced with relation to the background prevision. We can't define an electrical field of the form $E(r)\propto r^\alpha$, otherwise the redshift diverges for all points, if we consider materials \eqref{mm}. The only type of materials that allow such configuration are given by
\begin{equation}
    \epsilon(E) = \varepsilon + \sum_n c_n E^{-n}\;. \label{mm2}
\end{equation}
However, they aren't of physical interest, because we want the electromagnetic field to vanish at the infinity. We are going to focus our analyzes in materials \eqref{mm} and in the special class of materials $f_\pm=const.$ that makes the redshift remains the same as the background prevision.

\subsection{Optical horizons}
We don't expect the effective metrics \eqref{comp3} to produce alterations on the Killing horizons, because they define $\hat{g}_{00}^\pm = g_{00} f_\pm$, i.e., if $\hat{g}_{00}^\pm (r_h) = 0$ we obtain imediatly $g_{00}(r_h)=0$, which predicts the same positions as the background metric. One could argue that it is also possible the existence of optical horizons given by $f_\pm(r_h) =0$, however by definitions \eqref{defmeio} we see that this is only true for $\epsilon(E) \longrightarrow \infty$. Given \eqref{mm}, the only solution for this preposition is $r_h=0$ (which is a spacetime singularity, that we know the effective metric doesn't alter). On the other hand, for the special case were $f_+=const.$ we could obtain $r_h=\infty$, which seems unexpected, but its only implication is that light is trapped in a region because of the dielectric medium, however, this region includes all the universe. It is possible that this solution reflect the fact that (in this example) light can only propagate inward, i.e., that it has a negative velocity, associated to a wave vector $k$ with negative radial component. We conclude that the optical horizons coincide with the Killing horizons of the background metric.

\section{Special cases}\label{sec4}

\subsection{The class of material mediums with $f=const.$}\label{f1}
A particular case of interest occurs when $f_\pm (r)=const.$, since it gives a limit situation which doesn't alter the geometrical redshift with relation to the background predictions. We note that it is possible to have two cases: I- $f_-=f_+=const.$, which represents Gordon case, or II- $f_- \neq const.$, $f_+=const.$ . We show bellow their consequences. 

\subsubsection{The Gordon case}\label{GGordon}
By considering relation
\begin{equation}
    \frac{1}{\mu \epsilon(E)} = const.\;,
\end{equation}
one obtain
\begin{equation}
    \epsilon(E) = \varepsilon\;.
\end{equation}
Which implies that for this permittivity the geometrical redshift and the gravitational deflection of light coincide with the predictions of the background gravitational field, for both polarizations. The effective potential predicts corrections, given by equation \eqref{efpot}. The effective potential for both polarizations must be equal, so they follow the same path and the effects of polarization vanish. As the permittivity is constant, we have that $\xi=0$ and $f_\pm=f=\frac{1}{\mu\epsilon}$ , with  $\epsilon=const.$ and $\mu=const.$ . This makes the components of the effective metric simplify to
\begin{eqnarray}
    \hat{g}_{00}^\pm &=& \Delta f\;,\nonumber\\
    \hat{g}_{11}^\pm &=& - \Delta^{-1} \;,
\end{eqnarray}
where $\Delta$ is given by Reissner-Nordstrom background. We note that the Gordon case is a particular situation where $f=const.$ and $\epsilon=const.$, which implies that  $\hat{g}_{00}^\pm (r_h) =0$ can only give $g_{00}(r_h)=0$, since $f=const.$, which means that we not expect new Killing horizons. The vacuum is a particular case of this example, represented by $\epsilon=\epsilon_0$ and $\mu=\mu_0$.

If the effective metric is given by Gordon case, we can compute its potentials $ V_\pm(r)= V (r)$, represented by
\begin{eqnarray}
    V (r)&=& \frac{ L_0^2}{r^2} \Delta\; + \left[1-n^2\right]  W_0^2 \label{gordonvef}\;,
\end{eqnarray}
where $n^2=\mu\epsilon$ is the square of the refractive index of the medium. Specially, if we consider that the material medium is the vacuum, we have $n=1$ and $V_{Gordon} =  V_{B}$. 

\subsubsection{The case $f_+=const.$}
Imposing the restriction $f_+(r)=const.$, we can obtain the most general permittivity $\epsilon(E)$ that satisfy 
 \begin{equation}
    \frac{1}{\mu\epsilon(1+\xi)}=const.\;.
\end{equation}
It is direct to integrate this relation, obtaining the result
\begin{equation}
     \epsilon (E) = \varepsilon + \frac{a}{E}\;.\label{159}
\end{equation}
This permittivity is reduced to the previous case only if $a=0$. In general, $f_+=const.$ with $f_-\neq const.$ . As a consequence, only the geometrical redshift for the positive polarization coincides with the background prediction. For both polarizations, we expect corrections on the gravitational deflection of light and on the effective potential. In fact, given the permittivity \eqref{159} we have
\begin{equation}
    f_-\;=\; \frac{1}{\mu(\varepsilon + a/E)}\;,\;\;\;\;f_+\;=\;\frac{1}{\mu\varepsilon}\;,\;\;\;\; (1+\xi)\;=\; \frac{\varepsilon}{\epsilon}\;,
\end{equation}
such that it is direct to substitute these relations in equations \eqref{efpot} and (\ref{gd}-\ref{gd2}) given an electrical field $E=E(r)$. We note that in this peculiar case $\frac{f_-^\infty}{f_-} = 0$, which implies that for a negative polarization the geometrical redshift must be null for all radius, while for a positive polarization it is equal to the background prevision.

\subsection{ A particular case for the permittivity} \label{4.2}
We expect by experimental data that the permitivitty $\epsilon(E) = \varepsilon + c_1 E + c_2 E^2$ is enough to approximately describe any dielectric medium. In fact, experimental data show us that for most materials $c_2$ is smaller than $c_1$ by many orders of magnitude\cite{photonics}. For that reason, when we are dealing with weak fields, it seems reasonable to expect Pockels effect to be more relevant than Kerr effect \cite{1977UsFiN.123..349B,PhysRevLett.88.063001,exp1}. Therefore, we are going to restric our analyzes to a permittivity given by
\begin{equation}
    \epsilon(E)= \varepsilon + c_1 E\;,\label{pockel1}
\end{equation}
with $E(r) = r_Q^2 / r^2$ representing the background electromagnetic field. We want to know how birefringence alter the propagation of light on a spherically symmetric curved spacetime. 

\begin{figure}[!ht]
    \centering
    \includegraphics[scale=0.7]{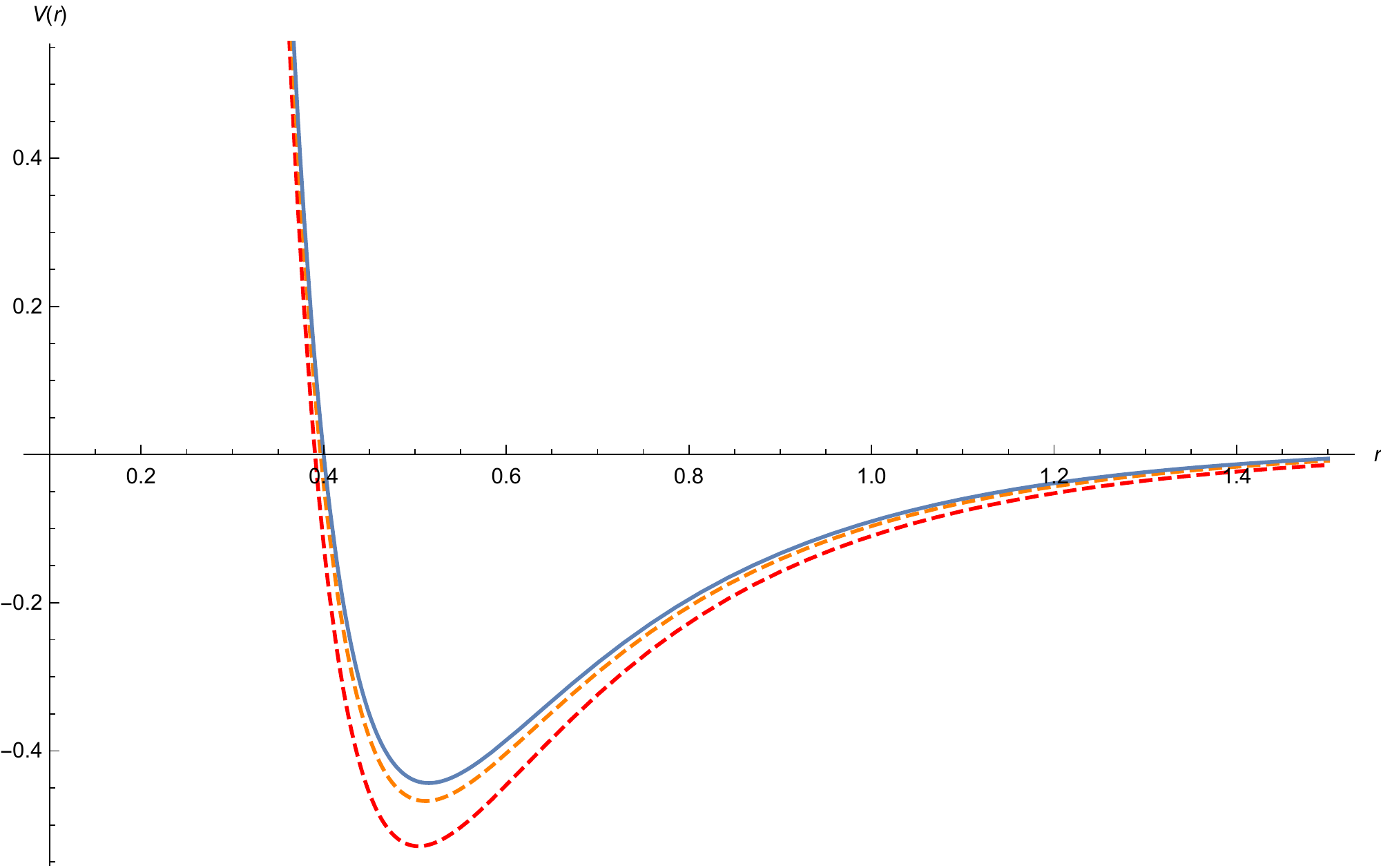}
    \caption{The effective potentials of example \eqref{pockel1}, given by equations \eqref{pockel2}, are represented by the dashed lines in this figure. The orange dashed line represents $V_-(r)$, while the red dashed line represents $V_+(r)$. In blue, we have what would be expected by considering only the background predictions. We see that in the presence of an electromagnetic field the potential represented by the blue line is separated in two potentials, one for each polarization, given by the dashed lines. For visualization, we considered $M>|Q|$ in the background. }
    \label{fig:my_label}
\end{figure}

By direct calculation, we obtain
\begin{equation}
    1+\xi \;=\; \frac{\varepsilon + 2c_1 E}{\epsilon}\;,\;\;\;\;f_+\;=\; \frac{1}{\mu (\varepsilon + 2c_1 E)}\;,\;\;\;\;f_-\;=\;\frac{1}{ \mu (\varepsilon + c_1 E)}\;.
\end{equation}
The effective potentials are given by
\begin{eqnarray}
       V_+(r) &=& \frac{L_0^2 \Delta}{r^2} \left[1+\frac{c_1 E}{\varepsilon + c_1 E}\right] + \left[1- \mu \frac{(\varepsilon + 2c_1 E)^2}{\varepsilon + c_1 E} \right] W_0^2\nonumber\\
       V_-(r)&=& \frac{L_0^2 \Delta}{r^2} + \left[1 - \mu (\varepsilon + c_1 E)\right] W_0^2\;, \label{pockel2}
\end{eqnarray}
for each polarization. We show in figure \ref{fig:my_label} how the presence of an electromagnetic field alter the potential of the background.

For the deflection of light, we can not solve the integrals (\ref{gd}-\ref{gd2}) analitically, but we can consider a weak field approximation if we expand functions $\Phi_\pm (r)$ with the aim to evaluate these integrals at a given order. We show that by this method these integrals can be solved analitically and they give corrections to GR predictions. In order to do that, we consider   $\frac{2m}{r}\;\sim\; \frac{r_q^2}{r^2} \ll 1$ and $\frac{2m}{r_0}\;\sim\; \frac{r_q^2}{r^2_0} \ll 1$ togheter with condition $\frac{c_1}{\varepsilon}\ll 1$, which is expected for most materials. Then, we can simplify
\begin{eqnarray}
       \frac{\Delta(r_0)}{\Delta(r)} \approx 1 + 2m \left(\frac{1}{r}-\frac{1}{r_0}\right) - r_q^2 \left(\frac{1}{r^2}-\frac{1}{r_0^2}\right)\;.
\end{eqnarray}
For $c_1^+ = 2 c_1^-$ and $c_1^- = c_1$, we have
\begin{equation}
     \frac{f_\pm (r_0)}{f_\pm (r)} \approx 1 + \frac{c_\pm r_q^2}{\varepsilon} \left(\frac{1}{r^2} -  \frac{1}{r_0^2}\right)\;,
\end{equation}
such that if we define $a_\pm = 1 - \frac{c_\pm r_q^2}{r_0^2 \varepsilon}$ we can isolate the term
\begin{eqnarray}
       \left(\frac{r}{r_0}\right)^2 \frac{\hat{g}^\pm_{00}(r_0)}{\hat{g}^\pm_{00}(r)} - 1 &\approx& \Bigg\{ 1 +  \frac{g(r)}{\left[\left(\frac{r}{r_0}\right)^2  - 1 \right] a_\pm}\Bigg\} \left[\left(\frac{r}{r_0}\right)^2  - 1 \right] a_\pm  \;, 
\end{eqnarray}
given the definition
\begin{eqnarray}
       g(r) &=& \left(\frac{r}{r_0}\right)^2 \left[ 2m \left(\frac{1}{r}-\frac{1}{r_0}\right) - r_q^2 \left(\frac{1}{r^2}-\frac{1}{r_0^2}\right)\right] \left[a_\pm + \frac{c_\pm r_q^2}{r^2 \varepsilon}\right]\;. 
\end{eqnarray}
It is possible to rewrite the expression
\begin{eqnarray}
       \frac{g(r)}{\left[\left(\frac{r}{r_0}\right)^2  - 1 \right] a_\pm} &=& \left[1 + \frac{c_\pm r_q^2}{a_\pm r^2 \varepsilon}\right] \frac{\left[ \frac{r_q^2 r}{r_0} \left(\frac{1}{r_0}+\frac{1}{r}\right) -\frac{2mr}{r_0}\right]}{\left(r  + r_0 \right)}\;,
\end{eqnarray}
and to verify that
\begin{equation}
    (1 + \xi)^{-1/2} \approx \left(1 + \frac{c_1 r_q^2}{r^2 \varepsilon}\right)^{-1/2} \approx  1 - \frac{c_+ r_q^2}{4r^2 \varepsilon}\;.
\end{equation}
As a consequence, we have
\begin{eqnarray}
       \Phi^+(r) &\approx& a_+^{-1/2} \frac{\Bigg\{1 + \frac{m}{r} - \frac{r_q^2}{2r^2}\Bigg\}}{\sqrt{\left(\frac{r}{r_0}\right)^2  - 1}}\Bigg\{1 + \left[1 + \frac{c_+ r_q^2}{a_+ r^2 \varepsilon}\right] \frac{\left[ \frac{mr}{r_0} - \frac{r_q^2 r}{2r_0} \left(\frac{1}{r_0}+\frac{1}{r}\right) \right]}{\left(r  + r_0 \right)}\Bigg\} \left(1 - \frac{c_+ r_q^2}{4r^2 \varepsilon}\right) \nonumber\;,\\
       \Phi^-(r) &\approx& a_-^{-1/2} \frac{\Bigg\{1 + \frac{m}{r} - \frac{r_q^2}{2r^2}\Bigg\}}{\sqrt{\left(\frac{r}{r_0}\right)^2  - 1}}\Bigg\{1 + \left[1 + \frac{c_- r_q^2}{a_- r^2 \varepsilon}\right] \frac{\left[ \frac{mr}{r_0} - \frac{r_q^2 r}{2r_0} \left(\frac{1}{r_0}+\frac{1}{r}\right) \right]}{\left(r  + r_0 \right)}\Bigg\}  \nonumber\;.
\end{eqnarray}
By making the substitution $u= r_0/r$ we can evaluate these integrals analitically. We obtain
\begin{eqnarray}
       \phi^+(u) - \phi^+_0 &=&
       \int_{0}^{u} \;\Phi^+(u)\; du\nonumber\\
       \phi^-(u) - \phi^-_0 &=&
       \int_{0}^{u} \Phi^-(u)\; du\;,\label{int}
\end{eqnarray}
where its functions are defined as
\begin{eqnarray}
       \Phi^+(u)&\approx&  a_+^{-1/2} \frac{\Bigg\{1 + \frac{mu}{r_0} - \frac{r_q^2 u^2}{2r^2_0}\Bigg\}}{\sqrt{1-u^2}}\Bigg\{1 + \left[1 + \frac{c_+ r_q^2 u^2}{a_+ r_0^2 \varepsilon}\right] \frac{\left[ \frac{m}{r_0} - \frac{r_q^2 }{2r_0^2} \right] - \frac{r_q^2 u }{2r_0^2}}{1+u}\Bigg\} \left(1 - \frac{c_+ r_q^2 u^2}{4r_0^2 \varepsilon}\right) \nonumber  \\
       \Phi^-(u) &\approx& a_-^{-1/2} \frac{\Bigg\{1 + \frac{mu}{r_0} - \frac{r_q^2 u^2}{2r^2_0}\Bigg\}}{\sqrt{1-u^2}}\Bigg\{1 + \left[1 + \frac{c_- r_q^2 u^2}{a_- r_0^2 \varepsilon}\right]  \frac{\left[ \frac{m}{r_0} - \frac{r_q^2 }{2r_0^2} \right] - \frac{r_q^2 u }{2r_0^2}}{1+u}\Bigg\}  \nonumber \;.
\end{eqnarray}
If we consider the product of terms $\frac{2mu}{r_0}\;,\; \frac{r_q^2 u^2}{r^2_0}\;,\;\frac{2m}{r_0}\;,\; \frac{r_q^2}{r^2_0}$ as approximately null, we can ignore the terms from its products. By considering $a_\pm^{-1/2} \approx 1 + \frac{c_\pm r_q^2}{2r_0^2 \varepsilon}$, we  arrive to
\begin{eqnarray}
       \Phi^+(u)&\approx&  h(u) + \frac{\frac{c_+ r_q^2}{2r_0^2 \varepsilon} - \frac{c_+ r_q^2 u^2}{4r_0^2 \varepsilon}}{\sqrt{1-u^2}}  \label{c1}  \\
       \Phi^-(u) &\approx& h(u) + \frac{\frac{c_- r_q^2}{2r_0^2 \varepsilon}}{\sqrt{1-u^2}}  \label{c2} \;,
\end{eqnarray}
where $h(u)$ defines the background prevision
\begin{equation}
    h(u)= \frac{1 + \frac{mu}{r_0} - \frac{r_q^2 u^2}{2r^2_0}}{\sqrt{1-u^2}} + \frac{ \frac{m}{r_0} - \frac{r_q^2 }{2r_0^2}  - \frac{r_q^2 u }{2r_0^2}}{(1+u)\sqrt{1-u^2}}\;.
\end{equation}
By solving the integrals \eqref{int} we find
\begin{equation}
    (\phi(r) - \phi_\infty)_\pm \approx arcsin\left(\frac{r_0}{r}\right) + \frac{m}{r_0} h_1(r) - \frac{r_q^2}{2r_0^2} h_2^\pm(r) \;, 
\end{equation}
with its functions being defined as
\begin{eqnarray}
 h_1(r) &=& 2 -\sqrt{1 - \left(\frac{r_0}{r}\right)^2} - \sqrt{\frac{1 - \frac{r_0}{r}}{1 + \frac{r_0}{r}}} \;, \nonumber\\
    h_2^+(r) &=& \left[\frac{3}{2} + \frac{c_+}{2\varepsilon}\right]arcsin\left(\frac{r_0}{r}\right)-\left[1 + \frac{c_+}{2\varepsilon}\right]\frac{r_0}{2r} \sqrt{1 - \left(\frac{r_0}{r}\right)^2}\\
     h_2^-(r) &=& \left[\frac{3}{2} + \frac{c_-}{\varepsilon}\right]arcsin\left(\frac{r_0}{r}\right)-\frac{r_0}{2r} \sqrt{1 - \left(\frac{r_0}{r}\right)^2}\;.
\end{eqnarray}
We know the light deflection in a region far from the source is given by \eqref{deflection}, which means that, as $h_1(r_0)=2$ and $h_2^\pm (r_0) =  \left[\frac{3}{2} + \frac{c_1}{\varepsilon}\right] \frac{\pi}{2}$, the polarizated photons have
\begin{equation}
     \Delta \phi_\pm \approx \frac{4m}{r_0} - \frac{r_q^2}{r_0^2} \left[\frac{3}{2} + \frac{c_1}{\varepsilon}\right] \frac{\pi}{2} \;. \label{gdv}
\end{equation}
Therefore, in first order both polarizations are deflected by the same angle, so in this approximation they are indistinguable. Both predicts the same corretion from background prevision. We expect high order terms to be negligible, by considering these approximations, but it is possible they are able to slightly distinguish both deflections. Nevertheless, due to their small order of magnitude, it probably wouldn't be possible to measure them phenomenologically. Therefore, for weak fields, we expect that a distant observer shouldn't be able to distinguish both polarizations by measuring only the gravitational deflection of light. However, the observer could determine the value of Pockels constant $c_1$, which is very important to predict properties of the dielectric medium. Different materials have different values for this constant\cite{photonics}, so if one could infer it at distance, it would be possible to compare it with experimental data known in literature and determine characteristics of the material. 

For the geometrical redshift, we find $f_\pm(\infty)=1/\mu\varepsilon$ and 
\begin{equation}
    \frac{f_+(\infty)}{f_+} = 1 + 2 \frac{c_1}{\varepsilon} E\;,\;\;\;\;\frac{f_-(\infty)}{f_-} = 1 + \frac{c_1}{\varepsilon} E\;.\nonumber
\end{equation}
One can notice that for $c_1>0$ the redshift has it intensity stronger and for $c_1<0$ it has it intensity smaller than the background prevision. The positive polarization is more affected than the negative polarization. Only at the infinity the predictions are equal to the background ($E(r)\longrightarrow 0$). One could make the same approximation for the redsfhit, considering weak fields, in order to find
\begin{equation}
    (1+z)_\pm \approx 1 + \frac{m}{r} + \frac{3m^2}{2r^2} - \frac{r_q^2}{2r^2}\left(1 - \frac{c_1^\pm}{\varepsilon}\right)\;.\label{reds}
\end{equation}
 We notice that by considering this limit, one is able to distinguish both polarizations. Therefore, relations (\ref{gdv}-\ref{reds}) could be used together to make a more efficient measure of Pockels constant $c_1$, for a material medium on a curved background. 

\subsection{Erasing the gravitational effect on photon propagation}
For certain geometries, like for instance, Reissner-Nordstrom metric, we can write the effective metric $\hat{g}^{\mu\nu}_-$ given in equation \eqref{ef} in a conformally flat form. In order to prove it, we are going to consider the isotropic line element
\begin{equation}
    ds^2 = A(r) dt^2 - B(r) (dr^2 + r^2 d\Omega^2),
\end{equation}
that reduces to Schwarzschild metric in the particular case where $A=(1-x)^2/(1+x)^2$ and $B=(1+x)^4$ with $x=m/2r$, after making the change $r=1/2 (R-m + \sqrt{R^2 -2mR})$. For an arbitrary dependence $A(r)$ and $B(r)$, we call the effective metric  of the negative polarization \eqref{ef} as $Q^{\mu\nu}$ . We find that it has components
\begin{eqnarray}
       Q^{00}&=& \frac{1}{A} + (\mu\epsilon -1)a^2 \;,\;\;\;\;Q^{11}\;=\;-\frac{1}{B}\;,\nonumber\\
       Q^{22}&=& - \frac{1}{B r^2}\;, \;\;\;\;\;\;\;\; \;\;\;\;\;\;\;\;\;\; Q^{33}\;=\; \frac{Q^{22}}{sen^2 \theta}\;,
\end{eqnarray}
for $v^\mu = a \delta^\mu_0$ normalized with $a= A^{-1/2}$. Thus, for the effective metric to be conformally flat $Q^{\mu\nu}= \xi \gamma^{\mu\nu}$, yields the values
\begin{eqnarray}
       \xi &=& \frac{1}{B}\nonumber\\
       \epsilon\mu &=& \frac{A}{B}\;. \label{440}
\end{eqnarray}
Note that the particular case of Schwarzschild background, in the exterior domain beyond the horizon, the value $\epsilon\mu$ from \eqref{440} is bounded $0<\epsilon\mu<1$.  As a consequence, there exists a conformal transformation that can eliminate the gravitational effects on  photon propagation. In other words, it is possible that the photon instead of moving in a geodesic on the geometry generated by gravity to follow a geodesic in the associated Minkowski geometry. This conformal transformation can only eliminate the gravitational effects felt by a photon with negative polarization.

\section{Conclusions}\label{FINAL}

In the present work, we have considered a static compact body surrounded by a nonlinear dielectric medium. We are motivated to analyze this configuration to study possible corrections to the propagation of light on an accretion disk with nonlinear electromagnetic properties.  We considered a simple model with a fixed background, restricting ourselves to a low energy situation. Usually, astrophysics consider perturbation theory to find corrections on the background due to the presence of matter, but it doesn't account $\mu,\epsilon\neq const$. So, as a future perspective, we could consider a perturbed background in order to account both corrections and construct a more realistic model. For the purpose of this article, a fixed background is sufficient to enumerate how a nonlinear medium can alter the propagation of light on a spherically symmetric curved spacetime. We list below our results,
\begin{itemize}
    \item  We show how a nonlinear dielectric medium alter the light deflection and the effective potential of light in such configuration.
    \item We present a formula describing the geometrical redshift modification, given a dependence $\mu=const.$, $\epsilon(E)$. We evaluate the effects of a general permittivity, increasing or decreasing the intensity of the redshift with relation to background predictions.
    \item The positions of killing horizons does not change, as consequence of the form of the effective metrics.
        \item In the case the background electromagnetic field polarize the material medium, different polarizations of light are distinguished by different corrections on the geometrical redshift and on the gravitational deflection of light. As a consequence, we have two possible paths for the trajectory of light in such configuration, that coincide when we turn off the electromagnetic field or if the permittivity is constant.
        \item We analyzed in section \ref{4.2} a particular case for the permittivity, showing its consequences on the propagation of light due to polarization. 
        \item For a negative polarization, we proved that in certain cases the effective metric is conformally flat.
            \item The analysis we present here predicts that in the absence of electromagnetic fields a nonlinear material medium would not produce measurable alterations in GR predictions. However, it shows that if an astrophysical body produces an electromagnetic field it can provoke measurable modifications in the propagation of light. These corrections are directly related to the permittivity of the medium $\epsilon(E)$. However, we don't expect a distant observer to suppose such permittivity dependence by knowning how the redshift and the gravitational deflection of light are altered in the presence of the field, since a observer far from the source would be dealing with weak fields and it is expected high order terms (associated to a nonlinear dependence on the electric field) to be negligible. Nevertheless, it is possible that the observer could determine Pockels constant $c_1$ by considering the corrections on the redshift and on the deflection angle. For this reason, by considering these corretions, an observer on earth should be able to determine properties of the medium at distance.
\end{itemize}
We remark that the effective metrics that represent $\mu=const.$, $\epsilon(E)$ and $\epsilon=const.$, $\mu(H)$ are related by a dual transformation, such that we expect to obtain equivalent results by considering $\epsilon=const.$, $\mu(H)$. We will analyze this case and consider other backgrounds in a future work.

A more realistic model could be constructed by considering Kerr-Newman background surrounded by mediums with $\mu=const.$, $\epsilon(E,H)$ and $\epsilon=const.$, $\mu(E,H)$ $-$ since we know these general cases can be described by an effective geometry. One could choose any observer because the system of equations allows this freedom. Besides it, there are some anisotropic formulations that permit the light propagation on a medium to be described by effective metrics. This suggests one could also consider working with anisotropic materials on a curved background, since they favor some direction for the polarization effect. If one is considering an accretion disk, it seems reasonable to consider a medium that surround the compact object. However, maybe different directions are favored, or the material is not equally distributed. As a future perspective, the model we present here could be generalized to a more arbitrary description.

\section*{Acknowledgements}

AG acknowledges the financial support of The Coordena\c c\~ao de Aperfei\c coamento de Pessoal de N\'ivel Superior - Brasil (CAPES) - Finance Code 001. MN acknowledges FAPERJ for a fellowship.
\appendix

\bibliographystyle{utphys2}
\bibliography{library}

\providecommand{\href}[2]{#2}\begingroup\raggedright\begin{thebibliography}{10}

\bibitem{lan84}
L.~D. Landau and E.~M. Lifshitz, {\em Electrodynamics of Continuous Media}.
\newblock Pergamon, New York, 1984.

\bibitem{T1}
W.~Mahmood and Q.~Zhao, ``The double jones birefringence in magneto-electric
  medium''. \href{http://dx.doi.org/10.1038/srep13963}{{\em Scientific Reports}
  {\bfseries 5} (09, 2015) }.

\bibitem{1984MolPh..52.1241G}
E.~B. {Graham} and R.~E. {Raab}, ``{A molecular theory of linear birefringence
  induced by crossed electric and magnetic fields}''.
  \href{http://dx.doi.org/10.1080/00268978400101911}{{\em Molecular Physics}
  {\bfseries 52} no.~5, (Jan., 1984) 1241--1249}.

\bibitem{1977UsFiN.123..349B}
N.~B. {Baranova}, I.~V. {Bogdanov}, and B.~I. {Zeldovich}, ``{New
  electro-optical and magneto-optical effects in liquids}''. {\em Uspekhi
  Fizicheskikh Nauk} {\bfseries 123} (Oct., 1977) 349--360.

\bibitem{Jones1948ANC}
R.~C. Jones, ``A new calculus for the treatment of optical systems. vii.
  properties of the n-matrices''. {\em Journal of the Optical Society of
  America} {\bfseries 38} (1948) 671--685.

\bibitem{PhysRevLett.85.4478}
T.~Roth and G.~L. J.~A. Rikken, ``Observation of magnetoelectric jones
  birefringence''. \href{http://dx.doi.org/10.1103/PhysRevLett.85.4478}{{\em
  Phys. Rev. Lett.} {\bfseries 85} (Nov, 2000) 4478--4481}.
  \url{https://link.aps.org/doi/10.1103/PhysRevLett.85.4478}.

\bibitem{PhysRevLett.88.063001}
T.~Roth and G.~L. J.~A. Rikken, ``Observation of magnetoelectric linear
  birefringence''. \href{http://dx.doi.org/10.1103/PhysRevLett.88.063001}{{\em
  Phys. Rev. Lett.} {\bfseries 88} (Jan, 2002) 063001}.
  \url{https://link.aps.org/doi/10.1103/PhysRevLett.88.063001}.

\bibitem{LlDXLAN}
J.~K. Ll.D., ``Xl. a new relation between electricity and light: Dielectrified
  media birefringent''. {\em Philosophical Magazine Series 1} {\bfseries 50}
  337--348.

\bibitem{exp1}
B.~Pelle, H.~Bitard, G.~Bailly, and C.~Robilliard, ``Observation of
  magneto-electric non-reciprocity in molecular nitrogen gas''.

\bibitem{bookex}
D.~Faccio, F.~Belgiorno, S.~Cacciatori, V.~Gorini, S.~Liberati, and
  U.~Moschella, {\em {Analogue Gravity Phenomenology: Analogue Spacetimes and
  Horizons, from Theory to Experiment}}.
\newblock Springer, Switzerland, 2013.

\bibitem{article}
P.~Piwnicki and U.~Leonhardt, ``Optics of moving media''.
  \href{http://dx.doi.org/10.1007/s003400000512}{{\em Applied Physics B-lasers
  and Optics - APPL PHYS B-LASERS OPT} {\bfseries 72} (01, 2001) 51--59}.

\bibitem{article1}
C.~Barceló, S.~Liberati, and M.~Visser, ``Analogue gravity''.
  \href{http://dx.doi.org/10.12942/lrr-2011-3}{{\em Living Reviews in
  Relativity} {\bfseries 8} (12, 2005) }.

\bibitem{article2}
C.~Barcel\'o, ``{Analogue black-hole horizons}''.
  \href{http://dx.doi.org/10.1038/s41567-018-0367-6}{{\em Nature Phys.}
  {\bfseries 15} no.~3, (2019) 210--213}.

\bibitem{article3}
M.~J. Jacquet, S.~Weinfurtner, and F.~K\"onig, ``{The next generation of
  analogue gravity experiments}''.
  \href{http://dx.doi.org/10.1098/rsta.2019.0239}{{\em Phil. Trans. Roy. Soc.
  Lond. A} {\bfseries 378} no.~2177, (2020) 20190239}.

\bibitem{article4}
J.~Muñoz~de Nova, K.~Golubkov, V.~Kolobov, and J.~Steinhauer, ``{Observation
  of thermal Hawking radiation and its temperature in an analogue black
  hole}''. \href{http://dx.doi.org/10.1038/S41586-019-1241-0}{{\em Nature}
  {\bfseries 569} no.~7758, (2019) 688--691}.

\bibitem{article5}
J.~Drori, Y.~Rosenberg, D.~Bermudez, Y.~Silberberg, and U.~Leonhardt,
  ``Observation of stimulated hawking radiation in an optical analogue''.
  \href{http://dx.doi.org/10.1103/PhysRevLett.122.010404}{{\em Phys. Rev.
  Lett.} {\bfseries 122} (Jan, 2019) 010404}.
  \url{https://link.aps.org/doi/10.1103/PhysRevLett.122.010404}.

\bibitem{article6}
T.~Torres, S.~Patrick, M.~Richartz, and S.~Weinfurtner, ``{Quasinormal Mode
  Oscillations in an Analogue Black Hole Experiment}''.
  \href{http://dx.doi.org/10.1103/PhysRevLett.125.011301}{{\em Phys. Rev.
  Lett.} {\bfseries 125} no.~1, (2020) 011301}.

\bibitem{article7}
S.~Eckel, A.~Kumar, T.~Jacobson, I.~B. Spielman, and G.~K. Campbell, ``{A
  rapidly expanding Bose-Einstein condensate: an expanding universe in the
  lab}''. \href{http://dx.doi.org/10.1103/PhysRevX.8.021021}{{\em Phys. Rev. X}
  {\bfseries 8} no.~2, (2018) 021021}.

\bibitem{article8}
T.~Torres, S.~Patrick, A.~Coutant, M.~Richartz, E.~W. Tedford, and
  S.~Weinfurtner, ``{Observation of superradiance in a vortex flow}''.
  \href{http://dx.doi.org/10.1038/nphys4151}{{\em Nature Phys.} {\bfseries 13}
  (2017) 833--836}.

\bibitem{article9}
J.~Steinhauer, M.~Abuzarli, T.~Aladjidi, T.~Bienaim\'e, C.~Piekarski, W.~Liu,
  E.~Giacobino, A.~Bramati, and Q.~Glorieux, ``{Analogue cosmological particle
  creation in an ultracold quantum fluid of light}''.
  \href{http://dx.doi.org/10.1038/s41467-022-30603-1}{{\em Nature Commun.}
  {\bfseries 13} (2022) 2890}.

\bibitem{Novello2003EffectiveG}
M.~Novello and S.~E.~P. Bergliaffa,
  \href{http://dx.doi.org/10.1063/1.1587103}{``Effective geometry''.}
\newblock 2003.

\bibitem{PhysRevD.78.045015}
V.~A. De~Lorenci and G.~P. Goulart, ``Magnetoelectric birefringence
  revisited''. \href{http://dx.doi.org/10.1103/PhysRevD.78.045015}{{\em Phys.
  Rev. D} {\bfseries 78} (Aug, 2008) 045015}.

\bibitem{LK}
V.~Lorenci and R.~Klippert, ``Non gravitational black holes''.
  \href{http://dx.doi.org/10.1590/S0103-97332004000700013}{{\em Brazilian
  Journal of Physics} {\bfseries 34} (12, 2004) 1367--1373}.

\bibitem{LM}
V.~Lorenci and M.~Souza, ``Electromagnetic wave propagation inside a material
  medium: An effective geometry interpretation''.
  \href{http://dx.doi.org/10.1016/S0370-2693(01)00588-3}{{\em Physics Letters
  B} {\bfseries 512} (02, 2001) }.

\bibitem{LKEM}
E.~Bittencourt, V.~Lorenci, R.~Klippert, M.~Novello, and J.~Salim, ``Analogue
  black holes for light rays in static dielectrics''.
  \href{http://dx.doi.org/10.1088/0264-9381/31/14/145007}{{\em Classical and
  Quantum Gravity} {\bfseries 31} (07, 2014) }.

\bibitem{PhysRevE.82.036605}
V.~A. De~Lorenci and D.~D. Pereira, ``Magnetoelectric birefringence as a unique
  effect in isotropic media''.
  \href{http://dx.doi.org/10.1103/PhysRevE.82.036605}{{\em Phys. Rev. E}
  {\bfseries 82} (Sep, 2010) 036605}.
  \url{https://link.aps.org/doi/10.1103/PhysRevE.82.036605}.

\bibitem{LK2}
V.~Lorenci and R.~Klippert, ``Analogue gravity from electrodynamics in
  nonlinear media''. \href{http://dx.doi.org/10.1103/PhysRevD.65.064027}{{\em
  Phys. Rev. D} {\bfseries 65} (02, 2002) }.

\bibitem{LK3}
V.~LORENCI and R.~Klippert, ``Electromagnetic light rays in local
  dielectrics''. \href{http://dx.doi.org/10.1016/j.physleta.2006.04.010}{{\em
  Physics Letters A} {\bfseries 357} (08, 2006) 61--65}.

\bibitem{LS}
V.~Lorenci and J.~Salim, ``Aspects of light propagation in anisotropic
  dielectric media''.
  \href{http://dx.doi.org/10.1016/j.physleta.2006.08.011}{{\em Physics Letters
  A} {\bfseries 360} (12, 2006) 10--13}.

\bibitem{DK}
D.~Pereira and R.~Klippert, ``Local nonlinear electrodynamics''.
  \href{http://dx.doi.org/10.1016/j.physleta.2010.08.033}{{\em Physics Letters
  A} (2010) }.

\bibitem{LKR}
V.~Lorenci, R.~Klippert, and D.~Teodoro, ``Birefringence in nonlinear
  anisotropic dielectric media''.
  \href{http://dx.doi.org/10.1103/PhysRevD.70.124035}{{\em Physical review D:
  Particles and fields} {\bfseries 70} (04, 2006) }.

\bibitem{Novello2000GeometricalAO}
M.~Novello, V.~A.~D. Lorenci, J.~M. Salim, and R.~Klippert, ``Geometrical
  aspects of light propagation in nonlinear electrodynamics''. {\em Physical
  Review D} {\bfseries 61} (2000) 045001.

\bibitem{Novello:2001gk}
M.~Novello and J.~M. Salim, ``{Effective electromagnetic geometry}''.
  \href{http://dx.doi.org/10.1103/PhysRevD.63.083511}{{\em Phys. Rev. D}
  {\bfseries 63} (2001) 083511}.

\bibitem{N}
M.Novello, ``Effective geometry in nonlinear electrodynamics''.
  \href{http://dx.doi.org/10.1142/S0217751X02013216}{{\em International Journal
  of Modern Physics A} {\bfseries 17} (01, 2012) }.

\bibitem{Lorenci2012TrirefringenceIN}
V.~A.~D. Lorenci and J.~P. Pereira, ``Trirefringence in nonlinear
  metamaterials''. \href{http://dx.doi.org/10.1103/PhysRevA.86.013801}{{\em
  Physical Review A} {\bfseries 86} (2012) }.

\bibitem{L}
V.~Lorenci, ``Effective geometry for light traveling in material media''.
  \href{http://dx.doi.org/10.1103/PhysRevE.65.026612}{{\em Physical review. E,
  Statistical, nonlinear, and soft matter physics} {\bfseries 65} (03, 2002)
  026612}.

\bibitem{LRY}
V.~Lorenci, R.~Klippert, and Y.~Obukhov, ``Optical black holes in moving
  dielectrics''. \href{http://dx.doi.org/10.1103/PhysRevD.68.061502}{{\em
  Physical Review D} {\bfseries 68} (10, 2002) }.

\bibitem{ME1}
M.~Novello and E.~Bittencourt, ``Metric relativity and the dynamical bridge:
  Highlights of riemannian geometry in physics''.
  \href{http://dx.doi.org/10.1007/s13538-015-0362-7}{{\em Brazilian Journal of
  Physics} {\bfseries 45} (08, 2015) }.

\bibitem{MSJVR}
M.~Novello, S.~Perez-Bergliaffa, J.~Salim, V.~Lorenci, and R.~Klippert,
  ``Analogue black holes in flowing dielectrics''.
  \href{http://dx.doi.org/10.1088/0264-9381/20/5/306}{{\em Classical and
  Quantum Gravity} {\bfseries 20} (02, 2003) 859}.

\bibitem{ABH}
M.~Novello, M.~Visser, and G.~Volovik, {\em {Artificial Black Holes}}.
\newblock World Scientific, Singapore; River Edge, U.S.A., 2002.

\bibitem{article10}
T.~G. Philbin, C.~Kuklewicz, S.~Robertson, S.~Hill, F.~Konig, and U.~Leonhardt,
  ``{Fiber-optical analogue of the event horizon}''.
  \href{http://dx.doi.org/10.1126/science.1153625}{{\em Science} {\bfseries
  319} (2008) 1367--1370}.

\bibitem{article11}
F.~Belgiorno, S.~L. Cacciatori, M.~Clerici, V.~Gorini, G.~Ortenzi, L.~Rizzi,
  E.~Rubino, V.~G. Sala, and D.~Faccio, ``{Hawking radiation from ultrashort
  laser pulse filaments}''.
  \href{http://dx.doi.org/10.1103/PhysRevLett.105.203901}{{\em Phys. Rev.
  Lett.} {\bfseries 105} (2010) 203901}.

\bibitem{article12}
F.~Belgiorno, S.~L. Cacciatori, G.~Ortenzi, L.~Rizzi, V.~Gorini, and D.~Faccio,
  ``Dielectric black holes induced by a refractive index perturbation and the
  hawking effect''. \href{http://dx.doi.org/10.1103/PhysRevD.83.024015}{{\em
  Phys. Rev. D} {\bfseries 83} (Jan, 2011) 024015}.
  \url{https://link.aps.org/doi/10.1103/PhysRevD.83.024015}.

\bibitem{article13}
F.~Belgiorno, S.~L. Cacciatori, G.~Ortenzi, V.~G. Sala, and D.~Faccio,
  ``Quantum radiation from superluminal refractive-index perturbations''.
  \href{http://dx.doi.org/10.1103/PhysRevLett.104.140403}{{\em Phys. Rev.
  Lett.} {\bfseries 104} (Apr, 2010) 140403}.
  \url{https://link.aps.org/doi/10.1103/PhysRevLett.104.140403}.

\bibitem{osti_4024998}
V.~I. Karpman, ``High-frequency electromagnetic field in plasma with negative
  dielectric constant.''.
  \href{http://dx.doi.org/10.1088/0032-1028/13/6/004}{{\em Plasma Phys. 13: No.
  6, 477-90(Jun 1971).} (1, 1971) }. \url{https://www.osti.gov/biblio/4024998}.

\bibitem{Perlick:2015vta}
V.~Perlick, O.~Y. Tsupko, and G.~S. Bisnovatyi-Kogan, ``{Influence of a plasma
  on the shadow of a spherically symmetric black hole}''.
  \href{http://dx.doi.org/10.1103/PhysRevD.92.104031}{{\em Phys. Rev. D}
  {\bfseries 92} no.~10, (2015) 104031}.

\bibitem{Wald:1984rg}
R.~M. Wald,
  \href{http://dx.doi.org/10.7208/chicago/9780226870373.001.0001}{{\em {General
  Relativity}}}.
\newblock Chicago Univ. Pr., Chicago, USA, 1984.

\bibitem{Carroll:2004st}
S.~M. Carroll, {\em {Spacetime and Geometry}}.
\newblock Cambridge University Press, 7, 2019.

\bibitem{Hadamard}
J.~Hadamard, {\em {Leçons sur la propagation des ondes et les équations de
  l'hydrodynamique}}.
\newblock A. Hermann, Paris, 1903.

\bibitem{Papapetrou:1975kj}
A.~Papapetrou, ``{Shock Waves in General Relativity}''.
  \href{http://dx.doi.org/10.1007/978-1-4684-0853-9_4}{{\em NATO Sci. Ser. B}
  {\bfseries 27} (1977) 83--101}.

\bibitem{gordon}
W.~Gordon, ``Zur lichtfortpflanzung nach der relativitätstheorie''.
  \href{http://dx.doi.org/10.1002/andp.19233772202}{{\em Annalen der Physik}
  {\bfseries 377} (03, 2006) 421 -- 456}.

\bibitem{NE}
M.~Novello and E.~Bittencourt, ``Gordon metric revisited''.
  \href{http://dx.doi.org/10.1103/PhysRevD.86.124024}{{\em Physical Review D}
  {\bfseries 86} (11, 2012) }.

\bibitem{photonics}
{\em Fundamentals of Photonics},
  \href{http://dx.doi.org/https://doi.org/10.1002/0471213748.ch18}{ch.~18,
  pp.~696--736}.
\newblock John Wiley and Sons, Ltd, 1991.

\bibitem{Weinberg:1972kfs}
S.~Weinberg, {\em {Gravitation and Cosmology}: {Principles and Applications of
  the General Theory of Relativity}}.
\newblock John Wiley and Sons, New York, 1972.

\bibitem{inverno:1992}
R.~D'Inverno, {\em {Introducing Einstein's Relativity}}.
\newblock Clarendon Press, 1992.

\end{thebibliography}\endgroup

\end{document}